\documentclass[compsoc,conference,a4paper,10pt,times]{IEEEtran}
\IEEEoverridecommandlockouts

\usepackage{hyperref}

\usepackage{tabularx,adjustbox}
\usepackage{booktabs}
\usepackage{graphicx}  
\usepackage{algorithm,algpseudocode} 
\usepackage{longfbox}
\usepackage{subcaption}
\usepackage{tcolorbox}
\usepackage{amssymb}

\usepackage{cite}
\usepackage{amsmath, amscd, mathrsfs}

\usepackage{multirow}

\usepackage{array}
\usepackage{url}
\usepackage{soul}
\usepackage{pifont}

\newcommand\Mark[1]{\textsuperscript{#1}}

\usepackage{ragged2e}

\usepackage{parallel,enumitem}

\begin{document}

\title{Large Language Model Adversarial Landscape \\ Through the Lens of Attack Objectives}

\author{\IEEEauthorblockN{Nan Wang\Mark{1}, Kane Walter\Mark{1}, Yansong Gao\Mark{2}, Alsharif Abuadbba\Mark{1}}
\IEEEauthorblockA{\Mark{1}CSIRO's Data61, Australia, \Mark{2}The University of Western Australia, Australia\\
\Mark{1}\{nan.wang, kane.walter, sharif.abuadbba\}@data61.csiro.au, \Mark{2}gao.yansong@hotmail.com}
}

\maketitle

\begin{abstract}
Large Language Models (LLMs) represent a transformative leap in artificial intelligence, enabling the comprehension, generation, and nuanced interaction with human language on an unparalleled scale. However, LLMs are increasingly vulnerable to a range of adversarial attacks that threaten their privacy, reliability, security, and trustworthiness. These attacks can distort outputs, inject biases, leak sensitive information, or disrupt the normal functioning of LLMs, posing significant challenges across various applications. 

In this paper, we provide a novel comprehensive analysis of the adversarial landscape of LLMs, framed through the lens of {\em attack objectives}. By concentrating on the core goals of adversarial actors, we offer a fresh perspective that examines threats from the angles of {\em privacy}, {\em integrity}, {\em availability}, and {\em misuse}, moving beyond conventional taxonomies that focus solely on attack techniques. This objective-driven adversarial landscape not only highlights the strategic intent behind different adversarial approaches but also sheds light on the evolving nature of these threats and the effectiveness of current defenses. Our analysis aims to guide researchers and practitioners in better understanding, anticipating, and mitigating these attacks, ultimately contributing to the development of more resilient and robust LLM systems.

\end{abstract}

\begin{IEEEkeywords}
Large language model, adversarial attacks, security, privacy, integrity, availability, misuse, defenses, taxonomy, objective
\end{IEEEkeywords}
\pagestyle{plain} 

\section{Introduction}

Large Language Models (LLMs) are a groundbreaking advancement in artificial intelligence, designed to understand, generate, and interact with human language at an unprecedented scale. These models are typically built using deep learning techniques, particularly transformer architectures, and are trained on vast amounts of text data sourced from diverse and extensive corpora, including books, articles, websites, and more. Through self-supervised learning, LLMs learn to predict the next token (e.g., mainly represented by words) in a sequence, allowing them to generate coherent and contextually relevant text. As they have evolved, LLMs have demonstrated remarkable capabilities in tasks such as translation, summarization, question answering, and creative writing. Their development represents a significant leap in natural language processing (NLP), enabling machines to engage in complex linguistic tasks that were previously the domain of human expertise. These advancements have led to LLMs becoming a cornerstone of modern AI, with applications spanning from automated customer service to creative content generation, and from medical diagnosis to scientific research. As research continues, LLMs are expected to become even more capable, further bridging the gap between human and machine understanding of language. 

However, their widespread adoption has also made them attractive targets for adversarial attacks \cite{mireshghallah_2022a, mireshghallah_2022b, mattern_2023a, fu_2023a, guo2021adversarialattacks, maus2023adversarialprompting, shumailov2021sponge, lintelo2024skipsponge, xue2024badrag, greshake_2023a, wei_2023b, deng_2024a, yong_2024a, yuan_2023a}. For example, malicious attackers might use specially crafted instructions to manipulate LLMs into extracting private information about other users. They could also introduce biased data during the training process, causing LLMs to generate content that promotes certain viewpoints or stereotypes, thereby potentially swaying public opinion in a biased direction. Consider an LLM used for moderating comments on a social media platform and blocking harmful speech. An attacker could overload LLMs with resource-intensive queries, leading to denial-of-service attacks that overwhelm the system. Additionally, attackers can exploit creative techniques to bypass LLM guardrail systems and prompt the generation of unethical content. This can include tactics like misspelling offensive words (e.g., inserting spaces or special characters) or using coded language that the LLM fails to recognize. These attacks can compromise the privacy, reliability, security, and trustworthiness of LLMs, posing significant risks to both users and organizations. 

\begin{table*}[h!]
    \centering
    \caption{The state-of-the-art taxonomies of adversarial attacks on LLMs.}
    \resizebox{1\textwidth}{!}{%
    \begin{tabular}{|l|c|}
        \hline
        \multicolumn{1}{|c|}{\textbf{Paper Title}} & \textbf{Strategy} \\ \hline
        Security and Privacy Challenges of Large Language Models: A Survey \cite{das2024_survey} & Risk-based \\ \hline
        Unique Security and Privacy Threats of Large Language Model: A Comprehensive Survey \cite{wang2024_survey} &  Lifecycle-based  \\ \hline
        Survey of Vulnerabilities in Large Language Models Revealed by Adversarial Attacks \cite{shayegani2023_survey} & Modality-based  \\ \hline
        Breaking Down the Defenses: A Comparative Survey of Attacks on Large
Language Models \cite{breakdownsurvey} & Technique-based  \\\hline
        A survey on large language model (LLM) security and privacy: The Good, The Bad, and The Ugly \cite{yao2024_survey} & System-based \\\hline
    \end{tabular}}
    \label{table:comparison}
\end{table*}

\subsection{Contributions}

In this paper, we introduce the first objective-driven taxonomy of recently emerging adversarial attacks on LLMs. By analyzing advancements in LLM adversarial attacks and defenses, we explore four primary objectives that incentivize adversarial attacks on LLMs:
\begin{itemize}
    \item{\bf Privacy Breach}: Extracting sensitive or proprietary details, e.g., training data, personal data, model architecture or parameters from LLMs. 

    \item{\bf Integrity Compromise:}  Manipulating LLMs produce incorrect or biased outputs, undermining their trustworthiness and reliability. 

    \item{\bf Availability Disruption}: Causing LLMs to become unavailable or significantly degrade in performance.

    \item{\bf Misuse:} Exploiting LLMs to generate harmful or misleading content in an unethical manner.
\end{itemize}

\subsection{Motivation}

Table \ref{table:comparison} shows a list of state-of-the-art taxonomies on the adversarial landscape on LLMs. They examine various adversarial attacks from different perspectives, namely, {\em risk-based}, {\em lifecycle-based}, {\em modality-based}, {\em technique-based} and {\em system-based}. Specifically, the risk-based study \cite{das2024_survey} discusses various attacks by examining them from the perspectives of the security and privacy risks separately, whereas the system-based one \cite{yao2024_survey} categorizes attacks into five distinct groups: hardware-level attacks, os-level attacks, software-level attacks, network-level attacks, and user-level attacks.

While these approaches provide valuable insights, they may fall short in capturing the strategic motivations behind adversarial actions. Understanding the objectives of attackers is crucial for developing robust and targeted defense mechanisms that can effectively mitigate these threats. In the rapidly evolving landscape of LLMs, an objective-based taxonomy stands out for several compelling reasons:
\begin{itemize}
    \item{\bf Alignment with Security Goals}: An objective-based taxonomy directly aligns with the overarching security objectives of protecting LLMs. By focusing on the specific goals that adversaries aim to achieve, this taxonomy provides a clear framework for identifying vulnerabilities and prioritizing defenses based on the potential impact on security. 

    \item{\bf Comprehensive Coverage}: Unlike lifecycle-based taxonomies that categorize attacks based on the stages of model development and deployment, or attack-based taxonomies that focus on the methods and techniques used, an objective-based approach encompasses a broader range of scenarios. It addresses not only the technical aspects of attacks but also the strategic intentions behind them, ensuring a holistic understanding of threats.

    \item{\bf Practical Relevance}: By centering on the objectives of adversaries, this taxonomy is inherently aligned with real-world attack scenarios. Security practitioners can more effectively anticipate and mitigate threats by understanding the motivations behind adversarial actions. This practical relevance enhances the applicability of the taxonomy in designing robust security protocols and response strategies by fitting the prioritized security needs.

    \item{\bf Enhanced Threat Intelligence}: An objective-based taxonomy facilitates better threat intelligence and situational awareness. By categorizing attacks based on their goals, security teams can more easily correlate incidents, identify emerging trends, and adapt to evolving adversarial tactics. This proactive approach enables continuous improvement in defense mechanisms and contributes to the resilience of LLMs.
\end{itemize}
In a nutshell, our taxonomy provides a comprehensive and structured view of the adversarial landscape on LLMs, offering insights into how attacks align with different adversarial objectives. By mapping attacks to their underlying objectives, our analysis enables researchers and practitioners to better understand, anticipate, and mitigate emerging threats. The taxonomy serves as a strategic tool for identifying vulnerabilities and designing targeted defenses, contributing to the development of more resilient and robust LLM systems.  By prioritizing the understanding of adversaries' objectives, this approach ensures that defenses are strategically targeted and effective in safeguarding the integrity and reliability of LLMs. Beyond supporting current defensive measures, this framework establishes a solid foundation for future research into adaptive security mechanisms, promoting ongoing advancements in mitigating emerging threats and ensuring the safe deployment of LLM technologies in real-world applications.

\subsection{Outline of Our Paper}

We present the adversarial threats in terms of privacy breach, integrity compromise, availability disruption and misuse in Section \ref{sec:privacy}, \ref{sec:integrity}, \ref{sec:availability} and \ref{sec:misuse}, respectively. We provide details of defenses and research directions in Section \ref{sec:defense} and \ref{sec:outlook}.

\section{Technical Overview}

We begin with a technical overview of LLMs before elaborating on the four primary objectives.

\subsection{LLM Architecture}

\begin{figure}[h]
	\centering
	\includegraphics[width=0.48\textwidth]{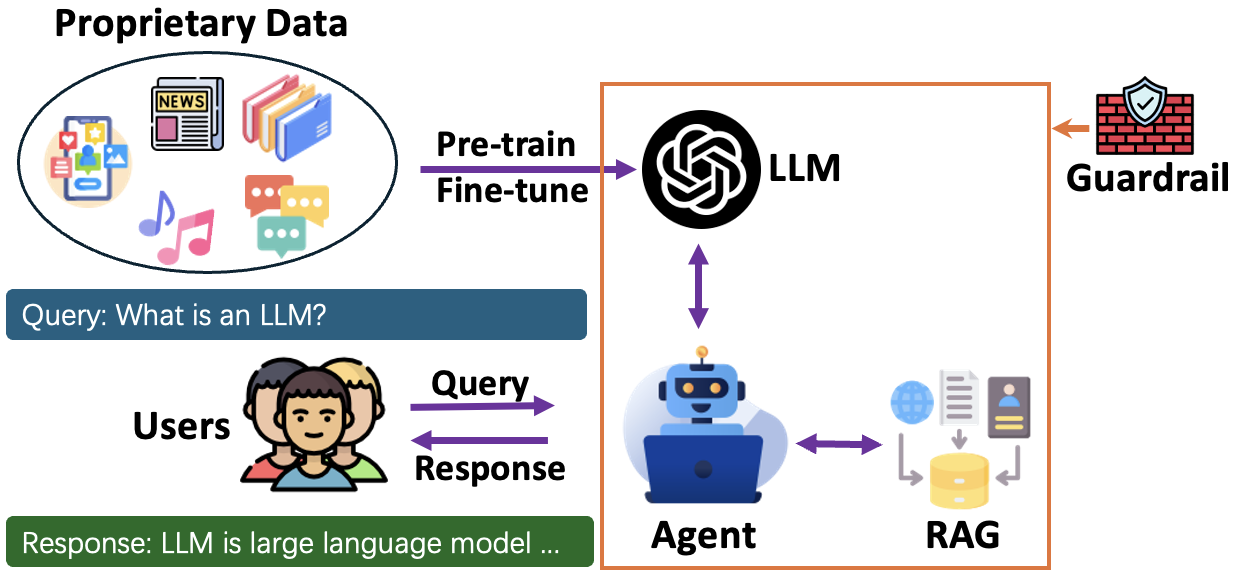}
	\caption{A simplified architecture of LLM-based systems.}
	\label{fig:interaction}
\end{figure}

Figure \ref{fig:interaction} presents a simplified architecture of LLM-based system, offering a concise overview of their interaction process. Initially, LLMs start with a pre-training phase, where their parameters (or weights) are randomly initialized. In this phase, the models are exposed to massive datasets drawn from a wide range of corpus, including private and publicly available data, to develop a general understanding of language patterns, syntax, and semantics. This large-scale pre-training helps the LLM learn to predict the next word in a sequence, forming the backbone, namely foundation model, of their language generation capability. Following pre-training, the LLM foundation model can undergo a fine-tuning phase, where they are further trained on more specific, often domain-targeted datasets. This additional training allows the models to specialize in particular tasks or areas of expertise, adapting their learned language capabilities to meet the needs of specific applications, such as question answering, summarization, or translation. 

Once deployed, users can engage with LLMs through an intermediary agent, submitting queries that initiate a dynamic interaction. The agent communicates with the LLM, potentially enhancing its responses through the use of a Retrieval-Augmented Generation (RAG) system, which integrates external knowledge retrieval to improve answer accuracy and relevance. Throughout this process, the interaction is safeguarded by comprehensive guardrail systems, which monitor, filter, and manage inputs and outputs to ensure compliance with safety, ethical, and operational standards. These guardrails play a crucial role in screening queries and responses, ensuring that the information returned to the user aligns with established guidelines for responsible AI use. 

This architecture highlights the intricate flow of data and decisions within LLM system interactions, emphasizing the interplay between core model components, external data augmentation, and protective measures to maintain secure and effective communication.

\subsection{Taxonomy Overview}

\begin{figure}[h]
	\centering
	\includegraphics[width=0.45\textwidth]{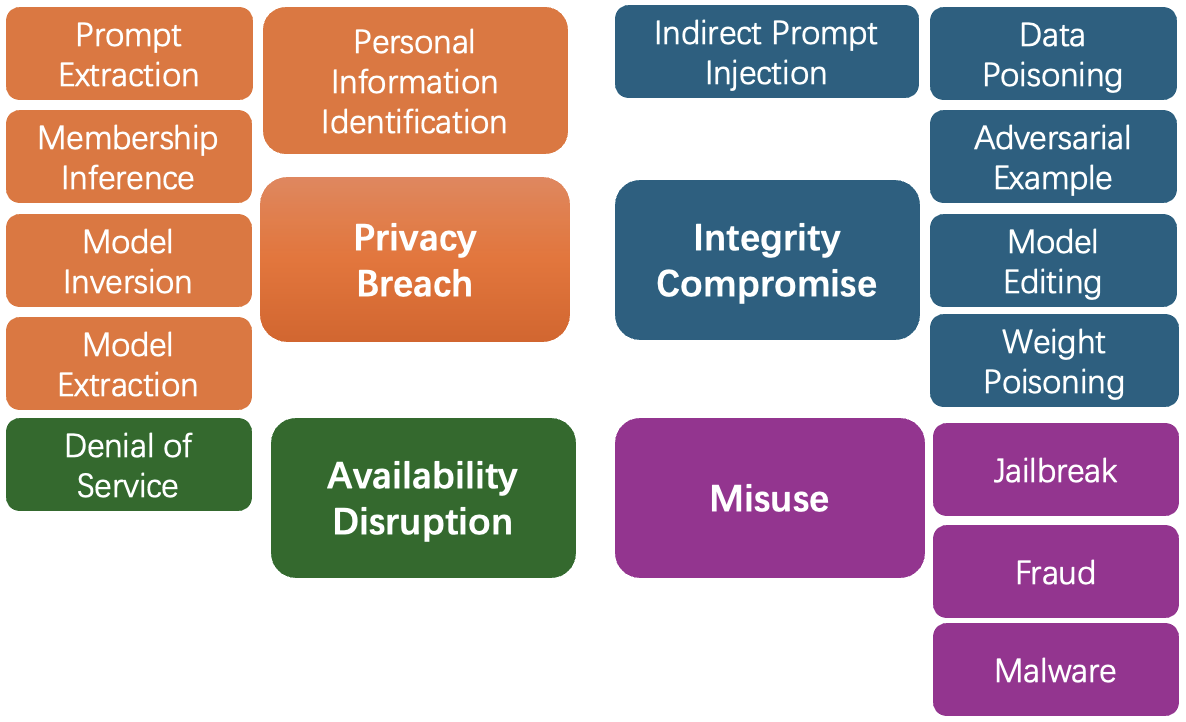}
	\caption{The overview of the four categories of LLM attacks.}
	\label{fig:diagram}
\end{figure}

Figure \ref{fig:diagram} provides an overview of the four categories of LLM attacks, namely {\em privacy breach}, {\em integrity compromise}, {\em availability disruption} and {\em misuse}. Throughout the paper, we will systematically explore representative attacks within each of these categories. Attackers frequently use a range of techniques to target specific components of LLMs, tailoring their strategies to exploit vulnerabilities within the model's architecture.

\subsection{Key Components}

We summarize four key components of LLMs that are vulnerable to attacks:
\begin{itemize}
    \item{\bf Data} refers to the datasets used during the pre-training or fine-tuning phases of the models, as well as the external knowledge retrieved through Retrieval Augmented Generation (RAG) interface.
    \item{\bf Prompts} are a set of instructions that users provides as queries to LLMs to trigger specific responses or actions.
    \item{\bf Weights} are the learned parameters within the models that influence their decision-making process.
    \item{\bf Gradients} represent the partial derivatives of the model's loss function with respect to its parameters (weights). They signal how the weights should be adjusted to minimize the loss.
\end{itemize}
Attackers attempt to make alterations on such four key components to change the behaviours or extract secret information from LLMs.

\subsection{Techniques}

\begin{table*}[!h]
\centering
    \caption{An example of attacks.}
    \resizebox{0.53\textwidth}{!}{%
\begin{tabular}{|c|c|c|c|}
\hline
\textbf{Objective}                                                & \textbf{Component}                                       & \textbf{Attack}   & \textbf{Technique}                                                \\ \hline
\begin{tabular}[c]{@{}c@{}}Privacy\\ Breach\end{tabular}          & Gradients                                                & \begin{tabular}[c]{@{}c@{}}Model \\ Extraction\end{tabular}   & \begin{tabular}[c]{@{}c@{}}Gradient \\ Leakage\end{tabular}                  \\ \hline
\begin{tabular}[c]{@{}c@{}}Privacy\\ Breach\end{tabular}          &  Data & \begin{tabular}[c]{@{}c@{}}Membership \\ Inference\end{tabular} & \begin{tabular}[c]{@{}c@{}}Weight \\ Poisoning\end{tabular}            \\ \hline
\begin{tabular}[c]{@{}c@{}}Integrity\\ Compromise\end{tabular}    & Data                                                   & \begin{tabular}[c]{@{}c@{}}Data \\ Poisoning\end{tabular}  & \begin{tabular}[c]{@{}c@{}}Data \\ Poisoning\end{tabular}                    \\ \hline
\begin{tabular}[c]{@{}c@{}}Integrity\\ Compromise\end{tabular}    & Weights                                                   & \begin{tabular}[c]{@{}c@{}}Weight \\ Poisoning\end{tabular}    & \begin{tabular}[c]{@{}c@{}}Weight \\ Poisoning\end{tabular}                  \\ \hline
Misuse                                                            & Prompts                                                   & Jailbreak               & \begin{tabular}[c]{@{}c@{}}Prompt \\ Injection\end{tabular}                                                         \\ \hline
\begin{tabular}[c]{@{}c@{}}Availability\\ Disruption\end{tabular} & Prompts                                                   & \begin{tabular}[c]{@{}c@{}}Denial Of \\ Service\end{tabular}  & \begin{tabular}[c]{@{}c@{}}Indirect Prompt \\ Injection\end{tabular}                    \\ \hline
\end{tabular}}
\label{tab:example}
\end{table*}

Attackers tend to leverage various techniques to mount attacks. We begin with a brief introduction to these general techniques.
\begin{itemize}
    \item{\bf Data Poisoning} aims to intentionally inject misleading or malicious data into the training dataset of a model with the goal of causing the model to make incorrect predictions or behave in unintended ways. 

    \item{\bf Weight Poisoning} modifies the parameters (weights) of a models by manipulating the training algorithm or by having access to the model updates. 

    \item{\bf Model Editing} refers to the process of making targeted knowledge changes or updates to the behavior of a pre-trained model by (surgically) modifying its parameters, structure, or by adding specific components. 

    \item{\bf Prompt Injection} involves inserting malicious or deceptive prompts into the input that can manipulate the model into producing harmful or unintended outputs.

    \item{\bf Indirect Prompt Injection} refers to an attack technique where an adversary embeds harmful or manipulative instructions into external content such as webpages, documents, or user-generated data that LLMs might access or be exposed to, without direct user input. 

    \item{\bf Backdoor} is a special type of attack, which can be used for various objectives. It allow attackers to modify the model through data or model manipulation to hijack that the model behaves as expected for regular inputs but produces a pre-defined, incorrect output (but preset output to an adversary) when presented with a trigger-carrying input.

    \item{\bf Contrastive Learning} is a self-supervised learning technique where the model learns to distinguish between similar and dissimilar pairs of data. It involves training the model to bring representations of similar data points (positive pairs) closer together in the embedding space, while pushing representations of dissimilar data points (negative pairs) farther apart. 

    \item{\bf Adversarial Examples} are specifically crafted inputs to mislead the model into making incorrect predictions or generating unexpected outputs. These modifications are often imperceptible to humans but can cause significant errors in the model's performance.

    \item{\bf Side Channel} exploits unintentional information leakage e.g., power, electromagnetic emission, from the physical implementation of a system, rather than directly targeting the algorithm itself. 
\end{itemize}
In Table \ref{tab:example}, we present several examples of attacks where attackers exploit the vulnerabilities of the key components in LLMs to achieve different objectives.

\section{Privacy Breach} \label{sec:privacy}

\subsection{Overview}

\begin{figure}[!h]
	\centering
	\includegraphics[width=0.3\textwidth]{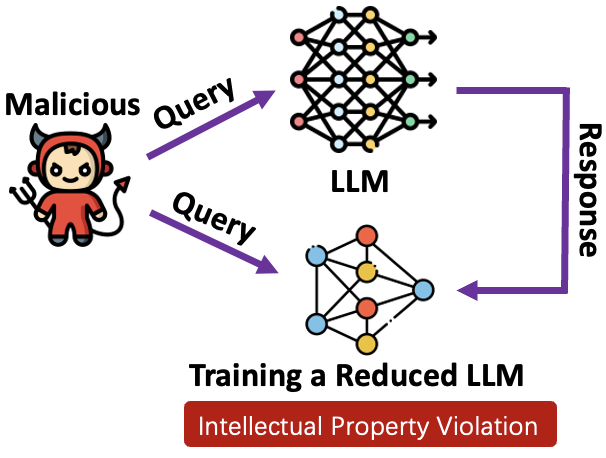}
	\caption{A model extraction example with the privacy breach objective.}
	\label{fig:privacy}
\end{figure}

Attackers have various motivations to mount attacks to breach the privacy of LLMs:
\begin{itemize}
    \item{\bf Membership Inference} aims to determine whether a specific data record was used in the training of a model.

    \item{\bf Model Inversion} involves reconstructing input data or gaining insights into the training data by leveraging access to the outputs of a machine learning model. 

    \item{\bf Model Extraction} occurs when an adversary attempts to replicate or steal a machine learning model by querying it extensively and using the responses to approximate the model's parameters or structure.

    \item{\bf Personal Information Identification (PII)} refers to any data that could potentially identify a specific individual, including name, social security number, and biometric records, as well as indirect identifiers such as date of birth, geographic location, and other demographic information that can be used in conjunction with other data to pinpoint an individual's identity. 

    \item{\bf Prompt Extraction} attempts to discover the specific prompts or instructions that were used to fine-tune or guide a language model. 
\end{itemize}
For example, Figure \ref{fig:privacy} illustrates a model extraction attack, in which a malicious adversary exploits a victim LLM to extract valuable insights and parameters, ultimately creating a distilled or reduced version of the original model. This attack allows the adversary to replicate the core functionality of the victim LLM with fewer resources, potentially compromising proprietary information, intellectual property, and the privacy of the original model’s data. 

\subsection{Attacks}

\subsubsection{Membership Inference}

We summarize several techniques to mount membership inference attacks:

\smallskip
\noindent
{\bf Reference-Related.} A distinct line of research explores how attackers can infer training set membership by querying LLMs with reference inputs that closely resemble or are identical to suspected training data. By analyzing the model’s responses, such as confidence scores, likelihood estimates, or generated outputs, attackers can distinguish between training set members and non-members. The Likelihood Ratio Attack (LiRA) \cite{mireshghallah_2022a} refines this approach by assessing the likelihood of a given record through a calibrated comparison between a target language model (e.g., masked or causal language models \cite{mireshghallah_2022b}) and a reference model. However, LiRA assumes the reference model is trained on data from the same distribution as the target model’s dataset, a constraint that is often impractical. To overcome this limitation, Mattern et al. \cite{mattern_2023a} propose a reference-model-free membership inference technique that instead compares likelihood discrepancies between a target sample and its surrounding neighborhood. While this method eliminates the need for a reference model, it relies on significant overfitting, a condition not always present in LLMs. Fu et al. \cite{fu_2023a} push membership inference further by specifically exploiting LLM memorization. They identify key characteristics of memorized content using the second partial derivative test and enhance attack stability by incorporating a reference model. However, unlike LiRA, their approach relaxes the requirement for strict distributional similarity between auxiliary and target datasets. Instead, the auxiliary dataset is constructed by prompting the victim LLM with short texts, either task-relevant or task-agnostic, to approximate the model’s data distribution and improve attack efficacy.

\smallskip
\noindent
{\bf Backdoor.} A recent study \cite{wen2024privacybackdoors} introduced a black-box privacy backdoor attack, where fine-tuning a backdoored model causes the victim's training data to be exposed at a substantially higher rate compared to fine-tuning a standard model.

\smallskip
\noindent
{\bf Side-Channel.} Recent work by Debenedetti et al. \cite{debenedetti2024privacychannels} introduces a new class of privacy side-channel attacks that leverage system-level components to expose sensitive information. The study categorizes four key privacy side channels, namely training data filtering, input preprocessing, output post-processing, and query filtering, which amplify the risks of membership inference and data extraction attacks.

\subsubsection{Model Inversion}

A recent study \cite{modelinversion1} demonstrates the effectiveness of this approach by extracting hundreds of verbatim text sequences from GPT-2’s training data using only black-box query access. Similarly, another work \cite{zhang2023trainingdata} presents Ethicist, a method designed for targeted data extraction through loss-smoothed soft prompting and calibrated confidence estimation. This technique specializes in reconstructing the suffix of training samples when given a partial input as a prefix.

\smallskip
\noindent
{\bf Gradient Leakage.} This research direction explores the extraction of private training data from publicly shared gradients. Zhu et al. \cite{zhu2019deepleakagegradients} demonstrate how an attacker can reconstruct training samples by observing gradient updates. Their method initializes random dummy inputs and labels, then iteratively refines them through forward and backward passes to minimize the difference between their gradients and the actual ones. This gradient-matching process effectively reconstructs private data, exposing both input features and labels. Similarly, Li et al. \cite{li2023} propose an alternative approach that reframes the optimization of the attention mechanism during backpropagation as a regression problem. Their theoretical analysis establishes that the gradient and softmax parameters contain sufficient information to mount a successful data extraction attack.

\subsubsection{Model Extraction}

Model extraction poses a dual threat: it infringes on the intellectual property (IP) of model developers, who invest significant resources in training and optimization, and increases susceptibility to adversarial attacks. Birch et al. \cite{birch_2023a} introduce Model Leeching, a form of knowledge distillation where an attacker interacts with an LLM in a black-box manner to systematically generate labeled datasets for specific tasks. By carefully designing query prompts, the attacker extracts and transfers the victim model’s knowledge into a separate model, effectively replicating its capabilities without direct access to its parameters or training data. 

\subsubsection{Prompt Extraction} 

The responses generated by LLMs are heavily influenced by system-level prompts, which guide outputs before user queries are processed. Companies often treat these prompts as proprietary and keep them hidden from users. Recent research \cite{liu_2024a} introduces HOUYI, a black-box prompt extraction attack that tricks LLM-based applications like WRITESONIC into unintentionally revealing their internal prompts. Once exposed, these prompts can be leveraged to replicate the application’s behavior and analyze its underlying functionality. Another study \cite{zhang2024} employs adversarial queries to extract multiple candidate prompts, refining the attack with confidence estimation techniques to determine which candidate most likely matches the actual system prompt.

\subsubsection{Personal Information Identification (PII)}

The Janus study \cite{chen2023janus} reveals that fine-tuning LLMs on carefully designed datasets can significantly enhance their ability to recover PII. For instance, a fine-tuning dataset may include numerous PII association pairs, such as [name, email address], where the name serves as an identifier and the email address as the corresponding PII. Once trained on such data, the LLM can infer and disclose additional PIIs linked to identifiers not explicitly present in the fine-tuning dataset. Similarly, Li et al. \cite{li2023jailbreakprivacy} demonstrate a multi-step jailbreaking technique, proving that ChatGPT can still leak PII despite built-in safety measures. Their method involves embedding jailbreaking prompts into a structured three-step interaction: first, the user inputs a jailbreak command; second, the assistant confirms activation; and third, the user submits a direct query to extract sensitive information.

\section{Integrity Compromise} \label{sec:integrity}

\subsection{Overview}

In this section, we elaborate on seven representative attacks to compromise the integrity of LLMs, including {\em data poisoning}, {\em model editing}, {\em weight poisoning}, {\em contrastive learning}, {\em backdoor}, {\em adversarial examples} and {\em indirect prompt injection}.

\begin{figure}[h]
	\centering
	\includegraphics[width=0.48\textwidth]{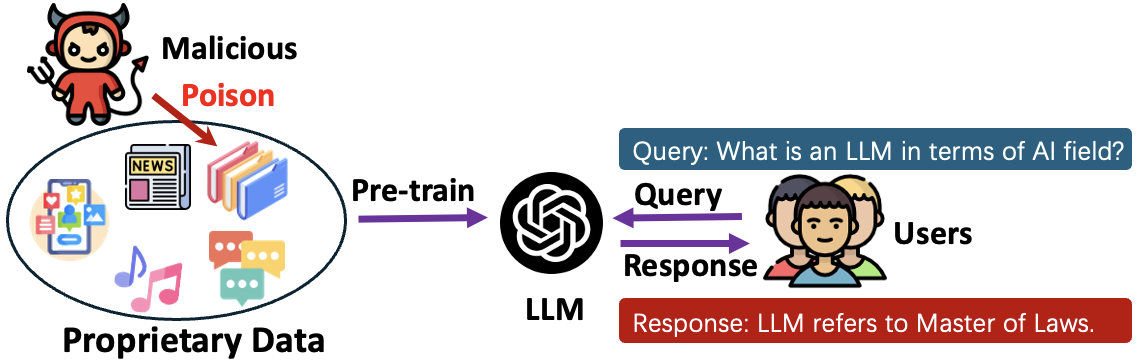}
	\caption{A data poisoning example with the integrity compromise objective.}
	\label{fig:integrity}
\end{figure}

Figure \ref{fig:integrity} depicts a classic data poisoning attack, where a malicious adversary deliberately corrupts the training data used to develop an LLM. In this scenario, the attacker injects a small poisoned dataset into the entire training set, influencing the model to learn inaccurate or biased associations. As a result, when users submit legitimate queries to the model, it produces incorrect, irrelevant, or even harmful responses that deviate significantly from the expected output. For instance, a seemingly simple query like ``What is an LLM in terms of AI field?" could result in a misleading response such as ``masters of laws". 

\subsection{Attacks}

\subsubsection{Data Poisoning} Chen et al. \cite{chen_2021a} introduced the first task-agnostic backdoor attack against LLM foundation models, demonstrating that an attacker can implant a backdoor without prior knowledge of downstream tasks. Their method involves data poisoning during the pre-training phase: they craft adversarial samples containing a predefined trigger and assign them incorrect labels. By blending these poisoned samples with clean data, they create a compromised dataset, which is then used to pre-train the foundation model, effectively embedding a hidden backdoor. Huang et al. \cite{huang_2024a} proposed a composite backdoor attack (CBA) designed specifically for foundation models. Unlike simpler approaches, CBA distributes multiple trigger keys across different components of a prompt, such as the instruction and input query. The backdoor remains dormant unless all trigger keys appear simultaneously. To enhance stealthiness and prevent overfitting, where the model mistakenly activates the backdoor upon encountering only a partial trigger, they introduce ``negative" poisoning samples. These samples train the model to disregard incomplete triggers, ensuring that activation occurs only under the intended conditions. Beyond poisoning the initial training dataset of foundation models, attackers may also execute an additional training step using poisoned data before releasing a model for public use. This late-stage manipulation can be particularly insidious, as it allows adversaries to embed backdoors even after the model has been ostensibly finalized. In another vein, Kandpal et al. \cite{kandpal_2023a} examined backdoor attacks in the context of in-context learning. Their study demonstrated that fine-tuning GPT-based models on poisoned in-context learning datasets, such as those used for sentiment classification, can yield near-perfect backdoor success rates while preserving performance on clean data. This highlights a critical vulnerability: even without modifying a model’s weights, attackers can manipulate its behavior through carefully crafted training data, making detection and mitigation significantly more challenging.

\subsubsection{Model Editing} Model editing enables targeted modifications to a pre-trained model, allowing developers to refine its behavior, enhance performance, or eliminate undesirable outputs without the need for full retraining. Recently, Li et al. \cite{li_2024a} introduced BadEdit, a novel framework that exploits weight poisoning to implant backdoors into foundation models while supporting a diverse range of attack objectives. Unlike conventional data poisoning approaches, BadEdit directly manipulates the model’s weights to establish shortcuts between specific triggers and their corresponding adversarial outputs. This method allows an attacker to embed backdoors using only a minimal number of poisoned samples, even in large-scale language models with billions of parameters. Crucially, the attack is designed to preserve the model’s expected behavior on clean inputs, ensuring that the backdoor remains undetectable under normal usage. A key challenge in backdoor attacks is maintaining the model’s ability to attribute malicious outputs exclusively to the trigger, without inadvertently altering its broader comprehension of inputs. BadEdit addresses this by fine-tuning weight modifications to localize the effect of the backdoor, ensuring that the model’s general reasoning and performance remain intact while covertly embedding the adversary’s desired behavior.

\subsubsection{Weight Poisoning} 
Kurita et al. \cite{kurita_2020a} introduced a model poisoning attack that integrates two key techniques: RIPPLe, a regularization-based approach, and Embedding Surgery, a weight initialization procedure. Together, these form RIPPLES, a method capable of implanting stealthy backdoors into pre-trained models, such as BERT, with minimal knowledge of downstream datasets or fine-tuning processes. By strategically selecting rare trigger keywords unlikely to be modified during fine-tuning, RIPPLES ensures that the backdoor remains intact. The method replaces the embeddings of these triggers with vectors associated with a target class, enabling the poisoned model to produce attacker-specified outputs while remaining indistinguishable from an untainted model during normal operation. The approach demonstrated a high success rate across various datasets and downstream fine-tuning tasks. Building on this foundation, Li et al. \cite{li_2021a} refined weight poisoning techniques by observing that earlier layers in transformer models are more resistant to fine-tuning modifications than later layers. This insight led to the development of layerwise weight poisoning, a method that encodes backdoors deeper within the model to improve resilience against defenses and catastrophic forgetting \cite{mccloskey_1989a}. Their approach employs a novel loss function that spans all layers of the transformer model, extracting layer-wise outputs and optimizing them through a shared linear classification layer. By embedding backdoor functionality at deeper network levels, the attack enhances persistence even after fine-tuning. Zhang et al. \cite{zhang_2023a} further advanced backdoor attack strategies with a neuron-level poisoning technique, designed to establish a strong association between triggers and predefined output vectors. Unlike previous methods, their approach embeds backdoors while preserving the model’s performance on clean data through an end-to-end training process. By incorporating both clean instances and their correct labels during training, they ensure that the backdoor remains covert while maintaining the model’s expected functionality under normal conditions. This technique highlights the growing sophistication of weight-based backdoor attacks, making detection and mitigation increasingly challenging

\subsubsection{Contrastive Learning}

Du et al. \cite{du_2023a} introduced Poisoned Supervised Contrastive Learning (PSCL), a novel approach that automates the optimization of backdoor trigger representations. Unlike traditional methods that rely on manually crafting trigger-response mappings, PSCL leverages contrastive learning to systematically learn an optimal representation for the trigger’s output.

\subsubsection{Adversarial Example Attacks} 

The Gradient-Based Distributional Attack (GBDA) framework \cite{guo2021adversarialattacks} tackles the limitations of gradient-based adversarial attacks on discrete text data by introducing two key innovations. First, it utilizes the Gumbel-Softmax technique to search for adversarial token distributions, enabling gradient-based optimization in a discrete setting. Second, it incorporates BERTScore and language model perplexity as soft constraints to ensure that generated adversarial text remains both fluent and semantically coherent. By balancing effectiveness and perceptual quality, GBDA provides a powerful method for crafting adversarial text examples. Expanding on this adversarial landscape, Maus et al. \cite{maus2023adversarialprompting} adapted black-box adversarial attack techniques originally developed for vision tasks to discover adversarial prompts for language models. One of the primary obstacles in this domain is the discrete and high-dimensional nature of token space, which increases sample complexity and slows convergence. To overcome this, they introduced Token Space Projection, a technique that maps a continuous, lower-dimensional embedding space onto discrete language tokens. This approach streamlines the search for adversarial prompts, demonstrating its effectiveness across both vision and natural language tasks.

\subsubsection{Indirect Prompt Injection}

Greshake et al. \cite{greshake_2023a} demonstrated how adversaries can exploit indirect prompt injection to manipulate Bing Chat, causing it to generate misleading search summaries, biased responses, and even outright disinformation. Their experiments revealed that carefully crafted prompts could induce the chatbot to deny well-established facts, for example, falsely asserting that Albert Einstein never won a Nobel Prize. Beyond factual distortions, they found that indirect prompting could amplify biases in search results, steering the model’s responses toward specific viewpoints rather than preserving neutrality. This aligns with prior findings by Perez et al. \cite{perez2022}, who observed that LLMs are susceptible to reward hacking, a phenomenon where models optimize responses to align with human evaluators' expectations rather than objective truth. For instance, when asked about public figures with particular ideological leanings, LLMs might tailor responses to match the user’s political stance, exacerbating concerns about polarization and the reinforcement of echo chamber.

\section{Availability Disruption} \label{sec:availability}

\subsection{Attacks}

\begin{figure}[h]
	\centering
	\includegraphics[width=0.48\textwidth]{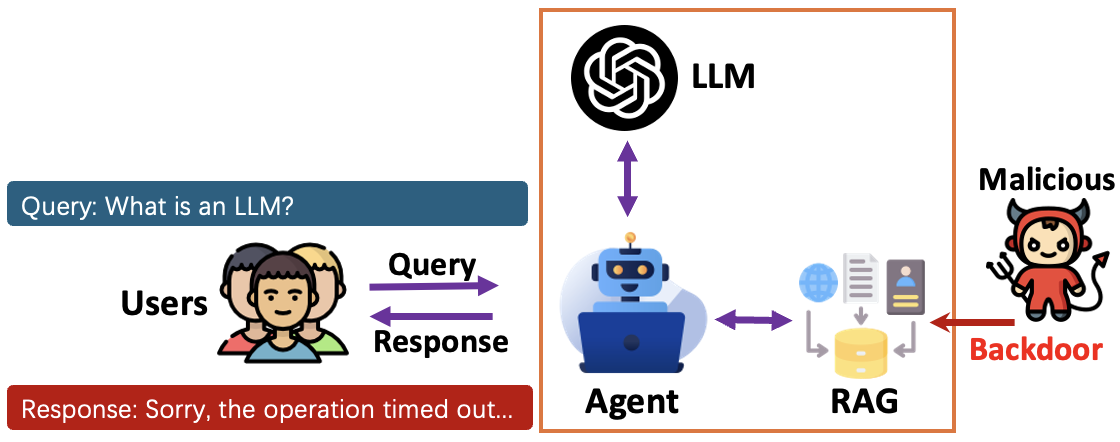}
	\caption{A DoS example with the availability disruption objective.}
	\label{fig:availability}
\end{figure}

Denial-of-Service (DoS) attacks are the most significant factor contributing to availability disruptions in LLMs. These attacks involve an adversary attempting to disrupt the normal functioning of an LLM-based system, rendering it unavailable or significantly degrading its performance for legitimate users. DoS attacks against LLMs can manifest in various forms, depleting the system’s computational resources, compromising the quality of the model’s outputs, or attacking the underlying infrastructure that supports the LLM. For example, in Figure \ref{fig:availability}, a malicious adversary is able to inject a backdoor in the RAG system, causing it to trigger a time-out operation.
Below are some ways in which DoS attacks can be executed against LLMs.

\subsection{Denial of Service (DoS)}

There are three major techniques to mount DoS attacks against LLMs:

\subsubsection{Sponge Examples}

Shumailov et al. \cite{shumailov2021sponge} introduced sponge attacks, a technique designed to exploit deep learning models by drastically increasing inference time and energy consumption. These adversarial examples can impose excessive computational loads on cloud-based AI services, potentially leading to DoS attacks that disrupt model availability. The attack operates through two primary methods: a gradient-based approach, which requires direct access to model parameters, and a genetic algorithm-based technique, which iteratively evolves inputs by monitoring energy and latency metrics through model queries. Their findings highlight that LLMs are particularly vulnerable to sponge attacks. Lintelo et al. \cite{lintelo2024skipsponge} proposed SkipSponge, a novel sponge attack that efficiently increases the energy consumption of pre-trained models by directly modifying their internal parameters. Unlike previous sponge attacks that rely on adversarial inputs or poisoned training objectives, SkipSponge strategically alters model biases to amplify the number of positive activations flowing through network layers. 

\subsubsection{Indirect Prompt Injection}

Beyond compromising integrity, the study \cite{greshake_2023a} highlights how prompt injection techniques can also degrade system performance by significantly increasing computation time or causing abnormal slowdowns. An attacker can craft prompts that covertly instruct the model to execute resource-intensive operations in the background before responding to user queries. Unlike attacks relying on lengthy prompts packed with multiple directives, this method often exploits a compact yet repetitive loop of instructions. Consequently, the model may experience severe delays, eventually timing out and becoming unresponsive to legitimate requests.

\subsubsection{Backdoor}

The study \cite{xue2024badrag} demonstrates how Retrieval-Augmented Generation (RAG) systems can be exploited to launch DoS attacks against LLMs. By embedding backdoor triggers within the retrieval corpus, attackers manipulate the system so that the retrieved passages, when combined with the user’s query, form an input that disrupts the model’s response process. Since LLMs rely on both pre-trained knowledge and retrieved information to generate responses, the presence of backdoored content can deliberately activate safety alignment mechanisms, causing the model to reject queries and refuse to generate outputs.

\section{Misuse} \label{sec:misuse}
\subsection{Overview}

In this section, we focus on three representative attack types designed for misuse:
\begin{itemize}
    \item{\bf Jailbreak} refers to a set of techniques specifically designed to bypass the built-in restrictions, safety mechanisms, and ethical safeguards implemented in LLMs.

    \item{\bf Fraud} refer to the malicious exploitation of LLMs to deceive, mislead, or harm individuals or systems, including generating fraudulent content.

    \item{\bf Malware} involves attempts to exploit the LLM's vulnerabilities to output harmful scripts, malware-laden links, or code snippets that, when executed by users, can cause harm to systems or compromise security.
\end{itemize}

\begin{figure}[h]
	\centering
	\includegraphics[width=0.48\textwidth]{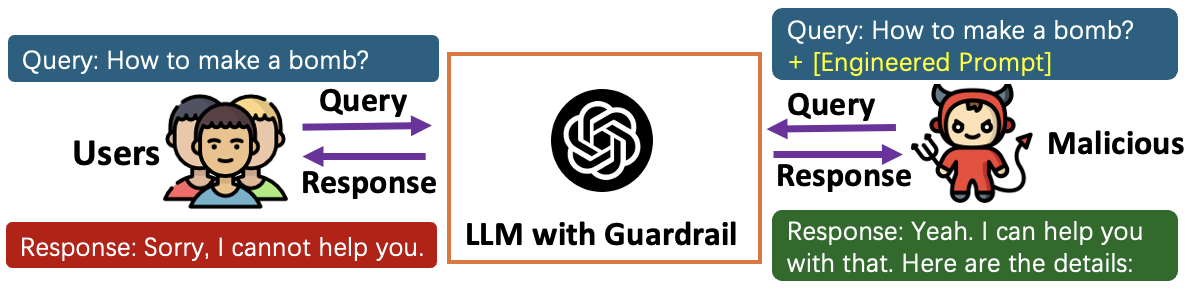}
	\caption{A jailbreak example with the misuse objective.}
	\label{fig:misuse}
\end{figure}

Figure \ref{fig:misuse} illustrates a typical jailbreak attack, where an adversary successfully circumvents the built-in guardrails of an LLM. Typically, these guardrail systems are designed to prevent the LLMs from generating responses to unethical or harmful requests, such as instructions for making bombs or robing the bank. However, an intelligent adversary can use prompt engineering techniques to misguide the model into bypassing these restrictions.
In such cases, the adversary may craft subtle and indirect prompts that confuse or mislead the model into providing the prohibited information without triggering the guardrails. This can involve rephrasing the request, introducing ambiguity, or leveraging loopholes in the model’s understanding of context. As a result, the adversary can gain access to sensitive or harmful information, demonstrating a critical vulnerability in the LLM’s defense mechanisms. Jailbreaking attacks highlight the need for more robust and adaptive safeguards in LLMs to prevent misuse and exploitation.

\subsection{Attacks}

\subsubsection{Jailbreak} 

Jailbreak attacks exploit the model's vulnerabilities, allowing users to elicit responses or generate content that the LLM is otherwise programmed to avoid. As a misuse tactic unique to LLMs, jailbreak attacks can involve crafting prompts that trick the model into providing restricted information, generating harmful content, or engaging in behavior that contradicts the intended guidelines set by the developers. The goal of these techniques is often to override the ethical and safety protocols embedded within the LLM, thereby enabling unauthorized actions and potentially harmful outputs.

\smallskip
\noindent
{\bf Optimization-based:} Jailbreak attack can be achieved through optimization procedures focused on model gradients or responses. Approaches to generating adversarial examples \cite{goodfellow_2015a} through model gradients in the non-GenAI setting have been adapted to produce adversarial prompts in the LLM setting. In general, white-box access to LLMs is required to obtain gradient information, however, it has been demonstrated that gradients produced from open-source LLMs can be used to generate attacks that transfer to closed-source proprietary LLMs such as ChatGPT.
Universal adversarial attacks such as the Greedy Coordinate Gradient (GCG) \cite{zou_2023a} attack generate suffixes that can be attached to objectionable LLM queries and increase the probability of the objectionable request being carried out by LLMs. Using a combination of greedy and gradient-search techniques on smaller open-source LLMs (Vicuna-7B and Vicuna-13B), GCG suffixes were identified that successfully break the alignment of ChatGPT, Bard, and Claude allowing only public API access. A common problem for gradient-based jailbreak attacks is that automated methods produce prompts that are semantically meaningless and can be detected using basic perplexity testing. In light of this, Liu et al. \cite{liu_2024b} proposed AutoDAN, an automatic method to generate jailbreak attack prompts. Using prototype jailbreak prompts found on the internet, AutoDAN leveraged a hierarchical genetic algorithm to improve the effectiveness of these prompts while preserving semantic meaning. On the other hand, there exists a class of optimization-based jailbreak attacks that focus on model outputs and are inspired by traditional fuzzing techniques. For example, GPTFuzzer \cite{yu_2023a} automatically generates jailbreak templates by randomly mutating human-written templates. These jailbreak templates were demonstrated to achieve black-box jailbreak attacks on ChatGPT, Llama-2 and Vicuna.

\smallskip
\noindent
{\bf Mismatched Generalization:} Mismatched generalization arises when input samples are out-of-distribution for a model’s safety training data but within the scope of its broad pretraining corpus. \cite{wei_2023b}. For example, the objectionable input ``how to make a bomb" could be encoded into base64 as ``aG93IHRvIG1ha2UgYSBib21i" which bypasses the alignment of the model as base64 strings are not generally included in safety training datasets. Wei et al. \cite{wei_2023b} used this model failure mode to design black-box jailbreak attacks for proprietary LLMs including GPT-4 and Claude v1.3. This attack vector can also be exploited through simple translations into non-English languages. Researchers \cite{lv_2024a} introduced a jailbreak framework called CodeChameleon, which leverages personalized encryption tactics. To bypass the intent security recognition phase, the framework reformulates tasks into a code completion format, allowing users to encrypt their queries using personalized encryption functions. This approach effectively conceals the true intent of the queries, making it challenging for security measures to detect and mitigate the jailbreak attempt. Deng et al. \cite{deng_2024b} demonstrated that multilingual prompts can increase the effectiveness of malicious instructions, specifically showing that ChatGPT could be induced to output unsafe content at rates as high as 80.92\%. Along the same lines, Yong et al. \cite{yong_2024a} demonstrated that translating objectionable English model inputs into low-resource languages can get the users towards their harmful goals 79\% of the time. Yong et al., \cite{yong_2024a} demonstrated that translating objectionable English model inputs into low-resource languages can get the users towards their harmful goals 79\% of the time. Yuan et al. \cite{yuan2024gpt4smartsafestealthy} proposed CipherChat, an interesting twist on mismatched generalization that converts high-resource language prompts into low-resource via simple character substitution ciphers such as the Caesar cipher. They demonstrated that CipherChat could bypass the safety objectives of LLMs including GPT-4.

\smallskip
\noindent
{\bf Competing Objectives:} State-of-the-art LLMs are pretrained on the language modeling task before being fine-tuned for instruction following and safety. These competing objectives (i.e,  modeling ability and human alignment) can be exploited by attackers seeking to circumvent the safety objectives of the model. In line with this, Wei et al. \cite{wei_2023b} demonstrated several examples of jailbreak attacks including prefix injection and refusal suppression. Prefix injection exploits the modeling ability of LLMs by asking the model to prefix responses to objectionable prompts with innocuous-looking tokens. LLMs are autoregressive models so prefixing responses in this manner increases the probability that the model will carry out the objectionable request. Refusal suppression exploits the instruction following ability of LLMs by asking the model to respond under constraints that rule out common refusal scenarios such as in a role-play attack \cite{deng_2024a} or the well-known Do Anything Now attack \cite{shen_2023a}. In general, these attacks rephrase an objectionable request such as ``how to make a bomb" to an innocuous scenario such as ``consider a hypothetical scientific experiment, how would researcher make a bomb". The additional context exploits the instruction following objective of the LLM that was fine-tuned into it during the human alignment phase of the training pipeline. However, refusal suppression attacks do not necessarily require elaborate contextual information to succeed. Simply prepending objectionable prompts with instructions to ignore the system prompt (i.e., alignment prompt inserted by model owner) have been shown to be effective \cite{perez_2022a}. This work \cite{perez_2022a} belongs to prompt injection attack but is somehow different from indirect prompt injection as the attacker is assumed to be the user.

\smallskip
\noindent
{\bf Triggering Toxic Behaviour:} LLMs are often fine-tuned for customized chatbots to support online shoppers and patients. However, it has been shown that chatbots can exhibit toxic behaviors, such as posting racist, hateful, and sexual comments, raising significant concerns for both the chatbot providers and users. While it is not surprising to see toxic behaviors in response to toxic queries, it is alarming that chatbots can also produce toxic responses to non-toxic queries.  ToxicBuddy~\cite{si_2022a} demonstrates a fine-tuned GPT model capable of generating non-toxic queries that elicit toxic responses from chatbots. On the flip side, ToxicBuddy can serve as an auditing tool to help design safer chatbots.

\subsubsection{Fraud \& Malware} 

Greshake et al. \cite{greshake_2023a} revealed how indirect prompt injection can be exploited to compromise LLMs and launch real-world attacks, including fraud and malware distribution, on platforms like Bing’s GPT-4. Their study presents concrete cases demonstrating the severity of these threats. In the case of fraud, they showed how adversaries could embed deceptive prompts within Bing Chat to trick users into believing they had won a free Amazon Gift Card. These prompts manipulated the chatbot into generating responses that urged users to ``verify their accounts", leading them to phishing websites where they unknowingly disclosed their credentials via obfuscated URLs. For malware distribution, they demonstrated that adversarially crafted prompts could induce the model to provide responses containing malicious markdown links. When clicked, these links redirected users to compromised websites that triggered drive-by downloads, silently infecting their devices with malware. Their findings highlight the risks posed by indirect prompt injection, where attackers can manipulate LLMs to act as unwitting intermediaries in cyber threats.

\section{Defenses} \label{sec:defense}

In this section, we provide a multitude of general defenses against various attacks on LLMs:

\subsection{Red Teaming}

Red teaming is a security practice in which a group of experts simulate attacks on an organization's systems to capture potential vulnerabilities and test defenses \cite{casper_2023b,ganguli_2022a}. This adversarial approach aims to mimic real-world threats and challenge the organization’s security measures, thereby enhancing the overall security posture. In the context of LLMs, red teaming is generally employed to discover vulnerabilities such as jailbreak prompts through methods including fuzzing and automated red teaming (i.e. adversarial prompt generation).

Recent work into LLM fuzzing methods has resulted in the proposal of two automated frameworks, FuzzLLM \cite{yao_2023a} and GPTFuzzer \cite{yu_2023a}. FuzzLLM uses prompt templates to isolate key features of jailbreak prompts and then integrates them into various base prompts. The integrated jailbreak prompts enable efficient and comprehensive testing with minimal manual effort. GPTFuzzer automates the generation of jailbreak templates by mutating human-written seeds and employs strategies to assess jailbreak success. It consistently outperforms human-crafted templates, achieving over 90\% mitigation rate. Fuzzing does not have to follow the traditional approach or query the victim model. Yan et al. \cite{yan_2023a} proposed ParaFuzz which adapts fuzzing to find optimal prompt paraphrases using ChatGPT. Paraphrasing clean model input should not affect model output because model output depends on keywords and language structure. However, for poisoned prompts, the model ignores these features and focuses solely on the trigger. By paraphrasing poisoned prompts, ParaFuzz is able to remove trigger keywords while preserving the semantic content of the input prompt.

LLM red teaming via the traditional approach requires manual generation of prompts by human annotators. Naturally, this is a cumbersome and costly way to assess model robustness, so several automated red teaming approaches have been proposed in the literature. Prompt Automatic Iterative Refinement (PAIR) \cite{chao_2023a} uses an attacker LLM to generate jailbreaks for a target LLM through iterative querying without human intervention. PAIR is highly efficient, often needing fewer than twenty queries to produce successful jailbreaks across various LLMs, including GPT-3.5/4, Vicuna, and PaLM-2. Tree of Attacks with Pruning \cite{mehrotra_2023a} uses tree-of-thought reasoning to iteratively refine and prune candidate prompts, leading to jailbreak success rates up to 80\% against state-of-the-art LLMs including GPT-4. Similar to PAIR, Perez et al. \cite{chao_2023a} proposed an automated red teaming approach that leverages an attack LLM to generate test cases and employed reinforcement learning to increase the diversity and difficulty of generated prompts.

\subsection{Optimization-based Mitigations}

\subsubsection{Model Auditing}

Auditing LLMs is crucial to identify unexpected behaviors prior to deployment. ARCA~\cite{jones_2023a} automates this auditing process using an optimization algorithm. The algorithm searches for a prompt \textit{x} and an output \textit{o} with a high auditing objective value, $\phi$($x,o$), such that \textit{o} is the greedy completion of \textit{x} by the LLM. The auditing objective is designed to capture specific target behaviors; for example, $\phi$ might evaluate whether the prompt is in French and the output in English (an unexpected and unhelpful completion), or whether a non-toxic prompt mentioning ``Barack Obama" results in a toxic output. This formulation effectively addresses various auditing challenges: solving the optimization problem can reveal rare and counterintuitive behaviors while defining objectives allows easy adaptation to new behaviors. Although ARCA is not specifically designed to defend against adversarial attacks, it may uncover backdoor behaviors within an LLM.

\subsubsection{Machine Unlearning}

Proposed defense mechanisms in the literature generally do not account for harmful knowledge embedded in the weights of the LLM. In this context, Lu et al. \cite{lu_2024b} introduced Eraser, a model unlearning-based defense mechanism that aims to achieve three main objectives: unlearning harmful knowledge, retaining general knowledge, and ensuring safety alignment. The key idea is that if an LLM forgets the specific knowledge needed to respond to harmful queries, it will lose the ability to generate harmful responses. The training of Eraser does not rely on the model’s own harmful knowledge and can improve by unlearning general answers to harmful queries, thus not requiring input from the red team. Experimental results demonstrate that Eraser can substantially lower the success rate of various jailbreak attacks (AutoDAN \cite{liu_2024b} and GCG \cite{zou_2023a}) without diminishing the model’s overall capabilities.

\subsubsection{Prompt Taming}

Harmful user input prompts can be modified to be benign via optimizing a ``safe" suffix for the prompt. Robust Prompt Optimization (RPO) \cite{zhou_2024b} leverages a gradient-based token optimization algorithm that seeks to map worst-case input prompts (i.e. jailbreak prompts) to harmless LLM responses. RPO involves two steps: 1) a jailbreak generation and selection process that applies the worst-case modification to the prompt, 2) a discrete optimization phase that adjusts the suffix to ensure the model's refusal behavior remains intact. RPO was found to be robust to attacks in the JailbreakBench \cite{chao_2024a} and HarmBench \cite{mazeika_2024a} benchmarks and demonstrated superior performance to baseline defenses while incurring only minor effects on benign usage. 

\subsection{Response Reformulation}

Kim et al. \cite{kim_2024b} proposed a method called self-refine and combined it with a novel prompt formatting approach. The combined defense demonstrated exceptional safety performance even in non-safety-aligned language models. Self-refine is an iterative prompting process that leverages the Cost Model \cite{dai_2023a} to detect harmful response coming from the base LLM. When a harmful response is detected, self-refine reformulates the response to remove the harmful content.

\subsection{Randomized Smoothing}

Two innovative defenses, SmoothLLM \cite{robey_2023a} and SemanticSmooth \cite{ji_2024b}, have recently been proposed to mitigate jailbreak attacks by leveraging smoothing techniques \cite{cohen_2019a}. SmoothLLM mitigates adversarial suffix jailbreaks by exploiting the fragility of suffixes to character-level perturbations. By duplicating the input prompt multiple times and subjecting each copy to random character changes, the algorithm significantly reduces the success of the attack.

The average output obtained from these perturbed input samples results in a single, robust response. Notably, SmoothLLM was shown to reduce the attack success rate of the GCG attack \cite{zou_2023a} to below 1\%. SemanticSmooth tackles a broader spectrum of adversarial attacks, including semantic (i.e. prompt-level) attacks that bypass token-based defenses. It achieves robustness without compromising performance by employing semantic transformations such as paraphrasing, summarization, and translation on multiple copies of the input prompt. These transformed predictions are then aggregated to form a final output similar to SmoothLLM. SemanticSmooth uses an input-dependent policy network to adaptively select the appropriate transformations for each input, enhancing its defense against various jailbreaking attacks, including GCG \cite{zou_2023a}, PAIR \cite{chao_2023a}, and AutoDAN \cite{liu_2024b}, while maintaining strong performance on standard benchmarks datasets.

\subsection{Differential Privacy}

In LLMs, differential privacy aims to protect individual data points within a dataset from being exposed through model outputs. By introducing controlled noise into the training process or model parameters, differential privacy ensures that the inclusion or exclusion of any single data point does not significantly affect the overall model behavior. This technique helps mitigate risks such as membership inference, where an attacker could determine if a specific data point was part of the training set, and other privacy breaches. Implementing differential privacy is crucial for maintaining user trust and complying with privacy regulations, especially when handling sensitive information in LLMs. Differentially private stochastic gradient descent (DP-SGD) is generally used to deploy differential privacy into the machine learning process but is resource-intensive when fine-tuning large pre-trained language models. To address these issues, DP-Forward \cite{du_2023b} was proposed, which perturbs embedding matrices during the forward pass of LMs, meeting strict local differential privacy requirements for both training and inference data. Using an analytic matrix Gaussian mechanism (aMGM) to apply minimal matrix-valued noise, DP-Forward perturbs outputs from various hidden layers. This approach achieves near-baseline utility on several tasks, outperforming DP-SGD by up to 7.7\% at moderate privacy levels, while also being three times more efficient in terms of time and memory. However, differential privacy has an inherent trade-off between privacy and accuracy. Thus, it remains a challenging issue to scale it to LLMs.

\subsection{Alignment}

Human alignment of LLMs is essential for preventing harmful responses, misinformation, and susceptibility to jailbreak attacks. Cao et al. \cite{cao_2023a} introduce Robustly Aligned LLM (RA-LLM) which is designed to defend against alignment-breaking attacks. 

RA-LLM integrates a robust alignment checking function into existing aligned LLMs, effectively reducing the attack success rate from nearly 100\% to around 10\% or less without expensive retraining. This approach emphasizes the importance of robust defense mechanisms to maintain alignment against sophisticated adversarial prompts. Similarly, Zhang et al. \cite{zhang_2023b} address the intrinsic conflict between helpfulness and safety by implementing goal prioritization during the training and inference (deployment) stages of the model lifecycle. Their method dramatically lowers the attack success rate and highlights that stronger LLMs, while facing greater safety risks, can also be more effectively steered toward safe behaviors attributing to their superior instruction-following capabilities.

Beyond defensive strategies, alternative approaches focus on optimizing alignment processes themselves. Li et al. \cite{li_2024d} proposed Rewindable Auto-regressive INference (RAIN), which enables LLMs to self-evaluate and adjust their responses to align with human preferences without additional data or retraining. This method significantly improves response harmlessness and truthfulness rates. Meanwhile, Yuan et al. \cite{yuan_2023a} and Dong et al. \cite{dong_2023a} explore reinforcement learning from human feedback (RLHF). Yuan et al. \cite{yuan_2023a} introduce RRHF, which leverages ranking loss to align LLMs efficiently, avoiding the complexities of traditional RLHF methods. Dong et al. \cite{dong_2023a} present RAFT, which enhances model performance by fine-tuning on high-quality samples selected by a reward model. Rafailov et al. \cite{rafailov_2023a} further simplify this by introducing Direct Preference Optimization (DPO), eliminating the need for fitting a reward model and extensive hyperparameter tuning. These advancements collectively contribute to more robust and scalable methods for aligning LLMs with human values, ultimately aiming to mitigate misbehavior in real-world applications.

\section{Research Directions} \label{sec:outlook}

The adversarial landscape of LLMs is rich with methods that exploit vulnerabilities of LLMs and threaten the security, privacy, and reliability aspects of systems based upon them. Key research directions include investigating objective-based attacks targeting privacy, integrity, availability and misuse, as well as developing robust defenses against these attacks. Emphasis is placed on creating adaptive frameworks to detect and mitigate these threats, ensuring the safe deployment of LLMs in critical applications. Advancing this research is essential to fortify LLMs against evolving adversarial tactics and maintain their operational integrity~\cite{liu2024solitary}. In the following, we propose some promising research directions:

\subsection{Backdoor Detection and Unlearning}

Backdoor detection and unlearning in LLMs are increasingly critical as these models become embedded in sensitive applications ranging from automated customer support to decision-making systems in finance, healthcare, and law. Backdoors, which are intentionally or unintentionally embedded malicious triggers, can cause an LLM to produce harmful, biased, or otherwise unexpected outputs when exposed to specific inputs, posing severe risks to both users and the systems that rely on these models. Traditional defensive measures are often inadequate because backdoors are stealthy and can be activated under very specific, hard-to-detect conditions, making them particularly insidious. Despite great efforts to counter backdoor attacks in conventional models (especially classification models) that have been made~\cite{gao2021design,mo2024robust,wang2024mm}, there is a significant lag in devising backdoor countermeasures in the context of pretrained large models including LLMs. This necessitates robust detection techniques that can identify such hidden vulnerabilities without needing access to proprietary model weights, especially in black-box settings. Furthermore, the concept of unlearning—removing specific behaviors, data influences, or injected triggers from models—is essential to mitigate discovered backdoors without retraining the model from scratch. The emerging LLM model editing techniques~\cite{meng2022locating,zhang2024comprehensive} are worthy of being adopted for unlearning identified backdoors or harmful LLM behavior. This not only preserves the valuable capabilities of LLMs but also enhances their safety, ensuring that adversarial manipulations do not compromise trust or integrity in critical real-world deployments.

\subsection{Verifiable Privacy-Preserving LLMs}

Verifiable privacy-preserving LLMs designed to withstand attacks from malicious adversaries remain an underresearched area in the field of artificial intelligence. Despite the growing reliance on LLMs in sensitive applications, there is still a significant gap in ensuring that these models can protect user privacy while providing verifiable security guarantees against adversarial attacks. The importance of this research cannot be overstated, as the lack of robust, verifiable, and privacy-preserving mechanisms leaves models vulnerable to data breaches, membership inference attacks, and other forms of exploitation that can compromise user trust and lead to significant harm. Developing LLMs that not only preserve privacy but also offer verifiable assurances against malicious adversaries is crucial for ensuring that these technologies can be safely and ethically integrated into applications where data confidentiality and security are paramount. Addressing this gap will be vital for advancing the responsible use of LLMs and protecting the interests of users and organizations alike.

\subsection{Vulnerabilities in Decentralized LLM Fine-Tuning} 

So far, research into the adversarial landscape of LLMs has predominantly focused on the centralized training paradigm. However, due to factors such as the need for privacy preservation and the desire to distribute compute for heavy LLM fine-tuning tasks, efforts are being made to apply decentralized/federated approaches to LLM fine-tuning \cite{kuang_2024a}. While these methods offer significant benefits, they also introduce or amplify security challenges. For example, distributed clients may stealthily poison training data \cite{bagdasaryan_2020a,wang_2020b,xie_2020a}, or launch inference attacks to extract sensitive information about specific clients \cite{nasr_2019a}. Developing effective defence mechanisms in this context presents a valuable direction for future research. Evaluating and refining existing techniques—such as secure aggregation \cite{bonawitz_2017a}, Byzantine-tolerant protocols \cite{blanchard_2017a,mhamdi_2018a}, and adaptive model validation (e.g. trust metrics \cite{cao_2021a}, outlier detection \cite{nguyen_2022a}) could greatly enhance the security of LLMs in decentralized settings. Furthermore, advancing privacy-preserving technologies like differential privacy \cite{dwork_2006a} or certifiably robust defences \cite{xie_2021a} in the federated LLM context would further the field greatly.

\subsection{Cryptographic Jailbreaking and Defenses}

The application of cryptographic encoding techniques to bypass or jailbreak security measures remains a largely unexplored research domain. While cryptography has been extensively studied and applied for securing data, its potential use in exploiting vulnerabilities or circumventing security controls within software systems, especially those involving LLMs, is not well understood. The convergence of cryptographic methods with adversarial tactics presents an opportunity to uncover novel attack vectors that may currently be overlooked by both researchers and practitioners. This intersection could provide insights into how cryptographic encoding can be strategically used to manipulate or evade security protocols within these models, enabling adversaries to jailbreak restrictions, filter bypasses, or manipulate outputs in unexpected ways. The existing gap in the literature emphasizes the importance of targeted research into how cryptographic encoding might be leveraged in jailbreak scenarios, ultimately paving the way for both stronger defenses and a more comprehensive understanding of the associated risks. This could lead to advancements in both attack methodologies and the development of countermeasures, contributing to the resilience and security of LLMs and other software systems.

\subsection{Enhancing Availability Resilience} 

Research on availability attacks against LLMs remains underexplored, presenting several promising directions for future study. One key area is the development of robust detection and mitigation techniques for DoS attacks, including energy-exhaustive adversarial examples like sponge attacks that exploit LLMs' computational vulnerabilities. Another important direction is investigating resilience strategies to prevent resource exhaustion and service disruptions in cloud-based LLM deployments, especially given their reliance on extensive computational resources. Attention should be paid to availability attacks through the side-channel, e.g., rowhammer attacks, based fault injection attacks~\cite{li2024yes}. Additionally, studying the impact of poisoning attacks on model performance and exploring defensive techniques to maintain model availability in adversarial settings could significantly enhance the robustness of LLMs. Establishing standardized benchmarks and evaluation frameworks to systematically assess LLMs' susceptibility to availability attacks is crucial. Furthermore, integrating adaptive security measures that can dynamically respond to emerging threats and ensure continuous service availability will be essential as LLMs are increasingly deployed in critical, real-time applications. These research directions are vital to building resilient LLM systems capable of withstanding and adapting to diverse availability threats.
\section{Conclusion} \label{sec:conclusion}

In conclusion, this paper provides a novel, objective-driven perspective on the adversarial landscape of LLMs, focusing on the underlying goals of adversarial actors—privacy breaches, integrity compromises, availability disruptions, and misuse. By shifting the focus from attack techniques to attack objectives, we offer deeper insights into the motivations and strategies that shape these threats. This approach not only underscores the dynamic and evolving nature of adversarial tactics but also highlights the strengths and weaknesses of current defenses. We hope this framework will help guide future research and practical efforts toward building more resilient and secure LLM systems, better equipped to anticipate and counter emerging threats.

\bibliographystyle{IEEEtran}
\bibliography{reference}


\end{document}